\documentclass[10pt,letterpaper]{article}

\usepackage[T1]{fontenc}
\usepackage[utf8]{inputenc}
\usepackage{newtxtext}
\usepackage{amsmath}
\usepackage{newtxmath}
\usepackage[margin=1in]{geometry}
\usepackage{microtype}
\usepackage{bm}
\usepackage{titlesec}
\usepackage{indentfirst}
\titleformat{\section}{\normalfont\large\bfseries}{\thesection}{1em}{}
\titleformat{\subsection}{\normalfont\normalsize\bfseries}{\thesubsection}{1em}{}
\titleformat{\subsubsection}{\normalfont\normalsize\bfseries}{\thesubsubsection}{1em}{}
\usepackage{graphicx}
\usepackage{tikz}
\usepackage{booktabs}
\usepackage[bottom]{footmisc}
\usepackage[numbers,sort&compress]{natbib}
\usepackage{caption}
\usepackage{float}
\usepackage{flafter}
\usepackage[section]{placeins}
\DeclareUnicodeCharacter{03B1}{\ensuremath{\alpha}}
\usepackage[hidelinks]{hyperref}
\usepackage{orcidlink}
\usepackage[nameinlink,noabbrev]{cleveref}
\graphicspath{{figures/}}

\newcommand{\CorrespondingEmail}{yuto.nakamura.t4@dc.tohoku.ac.jp}
\newcommand{\CorrespondingAuthorMark}{\textsuperscript{\normalfont\textdagger}}
\newcommand{\CorrespondingAuthorFootnote}{%
  \begingroup
  \renewcommand{\thefootnote}{\fnsymbol{footnote}}%
  \footnotetext[2]{Corresponding author: \href{mailto:\CorrespondingEmail}{\texttt{\CorrespondingEmail}}.}%
  \endgroup}

\newcommand{\FirstAuthorORCID}{0009-0008-4118-1078}
\newcommand{\SecondAuthorORCID}{}
\newcommand{\ThirdAuthorORCID}{}
\newcommand{\AuthorORCID}[1]{%
  \if\relax\detokenize\expandafter{#1}\relax
  \else\nobreak\hspace{0.3em}\orcidlink{#1}\fi}

\title{Phase-Controlled Time-Dependent Modal Analysis of Vortex-Gust Interaction with a Cylinder Wake}
\author{\textbf{Yuto Nakamura\textsuperscript{1}\CorrespondingAuthorMark{}\AuthorORCID{\FirstAuthorORCID}, Shintaro Sato\textsuperscript{2}\AuthorORCID{\SecondAuthorORCID}, and Naofumi Ohnishi\textsuperscript{2}\AuthorORCID{\ThirdAuthorORCID}}\\[6pt]
  \textsuperscript{1}Department of Aerospace Engineering, Nagoya University, Nagoya, Japan\\
  \textsuperscript{2}Department of Aerospace Engineering, Tohoku University, Sendai, Japan}
\date{} 

\newcommand{\TitleFontSize}{18}
\newcommand{\TitleLineSpacing}{22}
\newcommand{\LastCompiledDate}{%
  \number\year-\ifnum\month<10 0\fi\number\month-\ifnum\day<10 0\fi\number\day}
\newcommand{\ChrysanthemumLeaf}{%
  \path[draw=black,fill=black!5,line width=0.35pt,line join=round]
    (0,0) .. controls (3,3) and (2,6) .. (3,8)
    -- (5,6) -- (5,12) -- (8,10) -- (9,16) -- (12,13)
    -- (15,20) -- (18,13) -- (21,16) -- (22,10) -- (25,12)
    -- (25,6) -- (27,8) .. controls (28,4) and (23,0) .. (17,-1)
    .. controls (10,-3) and (5,-1) .. (0,0) -- cycle;
  \draw[line width=0.25pt] (0,0) .. controls (6,1) and (11,8) .. (15,18);
  \draw[line width=0.2pt] (5,2) -- (4,7) (8,5) -- (6,10)
    (11,9) -- (10,14) (6,2) -- (17,1) (10,7) -- (23,8)
    (13,12) -- (20,13);
}
\newcommand{\TitleOrnament}{%
  \begin{tikzpicture}[x=1pt,y=1pt,baseline=-2pt]
    \pgfmathsetlengthmacro{\ornamenthalfwidth}{0.5\linewidth}
    \path[use as bounding box] (-\ornamenthalfwidth,-25) rectangle (\ornamenthalfwidth,25);
    \draw[black!45,line width=0.25pt] (-\ornamenthalfwidth+8,0) -- (-17,0);
    \draw[black!45,line width=0.25pt] (17,0) -- (\ornamenthalfwidth-8,0);
    \foreach \side in {-1,1}{%
      \begin{scope}[xscale=\side]
        \draw[line width=0.5pt,domain=18:226,samples=100,smooth]
          plot (\x,{4.5*sin((\x-18)*4.5)});
        \draw[line width=0.35pt] (226,-2.65)
          .. controls (235,-2) and (235,7) .. (228,9)
          .. controls (222,11) and (220,5) .. (224,4);
        \foreach \pos/\angle/\scale in {35/-40/0.40,57/140/0.36,80/-30/0.46,103/145/0.40,126/-45/0.48,149/135/0.40,172/-35/0.44,195/145/0.34,215/-45/0.28}{%
          \pgfmathsetmacro{\vineheight}{4.5*sin((\pos-18)*4.5)}
          \begin{scope}[shift={(\pos,\vineheight)},rotate=\angle,scale=\scale]
            \draw[line width=0.6pt] (0,-3) -- (0,0);
            \ChrysanthemumLeaf
          \end{scope}
        }
        \draw[line width=0.3pt] (66,-3)
          .. controls (72,-14) and (88,-15) .. (89,-7)
          .. controls (90,-1) and (82,0) .. (81,-5);
        \draw[line width=0.3pt] (142,-1)
          .. controls (148,13) and (164,16) .. (166,9)
          .. controls (168,3) and (160,2) .. (159,7);
      \end{scope}
    }
    \foreach \angle in {0,22.5,...,337.5}{%
      \draw[line width=0.35pt,rotate=\angle]
        (0,1.4) .. controls (-1.8,4.4) and (-1.6,8.3) .. (0,9)
        .. controls (1.6,8.3) and (1.8,4.4) .. (0,1.4);
    }
    \foreach \angle in {11.25,33.75,...,348.75}{%
      \draw[line width=0.3pt,rotate=\angle]
        (0,1.2) .. controls (-0.9,3.0) and (-0.9,5.1) .. (0,5.7)
        .. controls (0.9,5.1) and (0.9,3.0) .. (0,1.2);
    }
    \fill (0,0) circle[radius=1.2pt];
    \fill (-19,0) circle[radius=0.7pt];
    \fill (19,0) circle[radius=0.7pt];
  \end{tikzpicture}\par}
\makeatletter
\renewcommand{\maketitle}{%
  \begin{center}
    \vspace*{8pt}
    \TitleOrnament
    \vspace{12pt}
    {\fontsize{\TitleFontSize}{\TitleLineSpacing}\selectfont\bfseries \@title\par}
    \vspace{14pt}
    \hrule height 0.45pt
    \vspace{16pt}
    {\normalsize\@author\par}
    \vspace{8pt}
    {\small \LastCompiledDate\par}
    \ifx\@date\@empty\else\vspace{10pt}{\normalsize\@date\par}\fi
    \vspace{8pt}
  \end{center}}
\renewenvironment{abstract}
  {\begin{center}\large\bfseries\textsc{Abstract}\end{center}\begin{quote}\normalsize}
  {\end{quote}\vspace{6pt}}
\makeatother

\begin{document}
\maketitle
\CorrespondingAuthorFootnote
\begin{abstract}
A vortex gust interacting with the periodic wake of a circular cylinder at $Re=100$ is studied using time-dependent dynamic mode decomposition with phase control (tDMDpc). To examine the phase dependence of the wake and lift responses, the same vortex gust is imposed at $64$ shedding phases. For these phase-controlled responses, proper orthogonal decomposition (POD) with time-independent spatial modes characterizes departures from the periodic wake, whereas tDMDpc resolves evolving structures and the frequency and growth rate of the fundamental mode. In this decomposition, the approaching gust is mainly captured by the zero-frequency mode. During the encounter, however, gust-related structures develop in the fundamental mode and persist near the cylinder as the gust convects downstream. The fundamental-mode coefficient magnitude is nearly uniform across phases at arrival but becomes phase dependent afterward. Consistent with this phase dependence, the near-wake vorticity amplitude of the fundamental conjugate pair decreases at $\alpha=\pi/2$ and increases at $\alpha=3\pi/2$ relative to the respective pre-encounter values, matching the inward and outward departures of the POD trajectories. To connect these structures to the cylinder lift, modal pressure and viscous stresses are integrated over the cylinder surface. The zero-frequency contribution carries most of the lift variation at arrival, whereas the fundamental conjugate pair dominates later oscillations. Phase averaging cannot separate these contributions because the fundamental oscillations no longer cancel across phases after the encounter. Their separation reveals that opposing contributions can yield nearly zero total lift despite an appreciable fundamental contribution. As the wake recovers, the lift amplitude approaches its undisturbed value, but a phase shift persists.
\end{abstract}

\section{Introduction}
Transient gust encounters can induce substantial variations in the flow around aircraft and the aerodynamic forces acting on them\cite{Fukami:2023,Fukami_2025_prf,taira2026extreme}. Predicting these responses requires understanding the interaction between the gust and the flow around the body, including variations in vortex shedding that persist after the gust has passed. Resolving the spatial development of the disturbance is therefore as important as predicting the time histories of lift and drag. Data-driven modal analysis\cite{POD1,POD15,POD0,DMD} provides low-dimensional descriptions of such transient fields and relates coherent structures to the aerodynamic response. However, transient gust encounters remain difficult to analyze because the gust convects downstream while the wake vortices deform and the amplitude of vortex shedding varies.

Gust models describe the velocity disturbance approaching a body in terms of its strength and variation in time\cite{Jones2020}. The velocity profile is used as an input for predicting the unsteady lift acting on the body during the encounter. Common examples include sharp-edged, sinusoidal (Sears)\cite{Sears1941,Atassi1984}, and $1-$cosine discrete gusts\cite{jones2022physics}. A sharp-edged gust imposes an abrupt onset of transverse velocity, a sinusoidal gust represents a periodic disturbance, and a $1-$cosine gust describes a smooth pulse of finite streamwise extent. These models represent the transverse velocity disturbance along the streamwise direction without explicitly resolving a localized vortex core.

Vortex gusts, in contrast, model coherent vortical disturbances such as vortices generated by upstream wings, rotors, or bluff bodies\cite{Rockwell1998}. The vortex gust is characterized by its radius $R$ and peak swirl velocity $u_{\theta,\max}$. The gust ratio is defined as $G=u_{\theta,\max}/U_\infty$, where $U_\infty$ is the freestream velocity. Recent studies have examined vortex-gust encounters with large $G$\cite{Fukami:2023,Fukami_2025_prf}. Numerical simulations have related the lift response to the vortex strength, size, and vertical position\cite{MartinezMurielFlores2020}, and experiments have shown that the vortex position and airfoil geometry affect separation and the lift peaks\cite{Qian2022}. The position of the gust normal to the freestream determines the extent of direct interaction between its core and the body or the shed vortices. Even on the same trajectory, the same gust meets a different arrangement of wake vortices depending on when it arrives. For a periodic wake, this timing is characterized by the phase of the vortex-shedding cycle. The wake of a circular cylinder is a simple periodic flow with a well-defined shedding phase, which can be varied while the gust radius, peak swirl velocity, and introduction position remain the same.

Phase-response analysis of periodic wakes quantifies the shift in vortex-shedding phase that remains after a weak velocity impulse is applied to the flow\cite{TairaNakao2018}. For vortex-gust--airfoil encounters, phase-amplitude reduction describes the phase shift and amplitude variation relative to the undisturbed periodic motion and relates these coordinates to the vorticity field around the airfoil and its transient lift coefficient\cite{FukamiNakaoTaira2024}. Relating particular flow structures to the aerodynamic forces also requires evaluating their individual contributions. Such force contributions have been evaluated for individual vortical structures\cite{LiWu2018ForceMap} and for individual flow modes\cite{PrakharSeoMittal2025MFP}. The evolving wake structures and the lift contributed by each frequency component must also be identified during the gust encounter.

Modal decompositions with time-independent spatial modes have been widely used to extract coherent structures from flow data\cite{POD1,POD15}. For example, proper orthogonal decomposition (POD)\cite{POD0,Berkooz1993} provides an optimal low-dimensional basis for a given dataset. A basis obtained from the undisturbed periodic wake consists of modes with time-independent spatial shapes. These modes provide a common reference for comparing disturbed flows, but their time-dependent coefficients contain contributions from both the passing gust and the deformation of the wake. Dynamic mode decomposition (DMD)\cite{DMD,Tu2014} decomposes time-resolved flow data into spatial modes, each with an oscillation frequency and a growth rate. In standard DMD, the spatial shape, frequency, and growth rate of each mode remain fixed over the analysis interval. Time-dependent modes allow the motion and deformation of the gust-related structures to be examined through their evolving spatial shapes.

Time-dependent modes have been used to analyze transient flows, such as those arising during vortex-gust interactions. Zhong et al.\cite{Zhong2025} track perturbation growth about the time-dependent flow using optimally time-dependent (OTD) modes\cite{BabaeeSapsis2016}. Zamani Ashtiani and Fukami\cite{ashtiani2025data} use time-dependent bases to represent turbulent airfoil wake--gust interactions. These approaches track the growth of perturbations or represent variations in the flow field during the gust encounter with a small number of modes.
In contrast, time-dependent dynamic mode decomposition with phase control (tDMDpc)\cite{mypaper_6} estimates a time-dependent linear operator from an ensemble of responses at different phases. Each mode has an instantaneous frequency and growth rate, and phase-specific coefficients quantify its contribution to the velocity field at each phase. Nakamura et al.\cite{mypaper_6} used tDMDpc to analyze the growth of vortex-shedding oscillations from a perturbed steady cylinder wake. The responses to the same gust at different shedding phases form an ensemble of velocity fields, suggesting that tDMDpc could be used to analyze the transient response of the wake.

In this study, we analyze the phase-dependent interaction between a Taylor vortex and the periodic wake of a circular cylinder at $Re=100$. We introduce the same gust at $64$ phases of the shedding cycle to form a phase-controlled ensemble. We use stationary POD to compare the disturbed wake with the undisturbed periodic state and tDMDpc to examine the evolving modal structures. We also apply tDMDpc to velocity and pressure and integrate the modal pressure and viscous stresses over the cylinder surface to evaluate the modal contributions to the lift.

\section{Modal analysis methodology}\label{sec:modal}
\subsection{POD}
POD extracts $r$ orthonormal basis vectors $\boldsymbol{\phi}_i(\boldsymbol{x})\ (i = 1, 2, \cdots, r) \in \mathbb{R}^{N}$ that optimally represent a given flow dataset $\boldsymbol{u}(\boldsymbol{x}, t_j)\ (j = 1, 2, \cdots, M) \in \mathbb{R}^{N}$, where $M$ is the number of snapshots in the dataset and $r < M$\cite{Berkooz1993}. This POD mode count $r$ is chosen independently of the SVD rank in Section~2.2. Here, $N$ denotes the number of degrees of freedom of the spatially discretized state vector. The inner product is defined as
\begin{equation}
\begin{split}
 \langle \,\boldsymbol{a}(\boldsymbol{x}), \boldsymbol{b}(\boldsymbol{x}) \rangle = \int_V \{\boldsymbol{a}(\boldsymbol{x})\}^{\top} \boldsymbol{b}(\boldsymbol{x}) d\boldsymbol{x},
 \end{split}
\end{equation}
where $\langle \cdot, \cdot \rangle$ denotes the inner product, $\boldsymbol{a}(\boldsymbol{x})$ and $\boldsymbol{b}(\boldsymbol{x})$ are $N$-dimensional vectors (bold symbols represent vectors), and the superscript $\top$ denotes transpose. In the inner product for complex fields, $\top$ is replaced by the conjugate transpose. The POD basis is given by
\begin{equation}
\begin{aligned}
 U&=[\boldsymbol{\phi}_1(\boldsymbol{x}),\,
       \boldsymbol{\phi}_2(\boldsymbol{x}),\,
       \cdots,\,\boldsymbol{\phi}_r(\boldsymbol{x})],\qquad U\in\mathbb{R}^{N\times r}.
\end{aligned}  
\end{equation}
which solves the following minimization problem:
\begin{equation}
\begin{aligned}
 U={}&\underset{U}{\operatorname{argmin}}\sum_{j=1}^{M}
       \Bigl\|\boldsymbol{u}(\boldsymbol{x},t_j)-\sum_{k=1}^{r}
      \langle\boldsymbol{u}(\boldsymbol{x},t_j),
      \boldsymbol{\phi}_k\rangle
      \boldsymbol{\phi}_k(\boldsymbol{x})\Bigr\|^2,\\
 &\text{subject to}\quad
   \langle\boldsymbol{\phi}_i,
   \boldsymbol{\phi}_j\rangle=\delta_{ij}.
\end{aligned}
\end{equation}
where the norm is induced by the inner product, $\|\boldsymbol{v}\|^2=\langle\boldsymbol{v},\boldsymbol{v}\rangle$. The POD modes are obtained by transforming the leading $r$ left singular vectors of the spatially weighted snapshot matrix back to physical space\cite{Sirovich1987}. The POD coefficient $a^s_k(t)$ of a flow field $\boldsymbol{u}(\boldsymbol{x}, t)$ is
\begin{equation}
a^s_k(t) = \langle \boldsymbol{u}(\boldsymbol{x}, t) , \boldsymbol{\phi}_k(\boldsymbol{x}) \rangle,
\end{equation}
where the superscript $s$ identifies the stationary POD coefficients.

\subsection{tDMDpc}
tDMDpc uses a time-dependent linear operator to capture the temporal evolution of eigenmodes in transient flows\cite{mypaper_6}. The original ensemble followed growth from phase-controlled perturbations of a steady wake; here, the ensemble consists of responses of a periodic wake to the same gust at different phases. We write the instantaneous linear approximation as
\begin{equation}
\frac{\partial \boldsymbol{u}(\boldsymbol{x},t)}{\partial t}\approx\mathcal{A}(t)\boldsymbol{u}(\boldsymbol{x},t).
\end{equation}
The operator is estimated by solving the DMD minimization problem
\begin{equation}
\underset{A(t)}{\text{argmin}} \, \left\|W^{1/2}\left[Y(t)-A(t)X(t)\right]\right\|_F,
\label{Aargmin_tDMDpc}
\end{equation}
where $\| \cdot \|_F$ denotes the Frobenius norm and $W$ is the diagonal matrix of cell-area weights. The snapshot-pair separation is $\Delta T$. If $\mathcal{A}(t)$ is assumed constant over this interval, it is related to the discrete-time operator by
\begin{equation}
A(t)\simeq\exp\!\left[\mathcal{A}(t)\Delta T\right].
\end{equation}
The datasets $X(t)$ and $Y(t)$ are constructed from realizations with different phases.
\begin{equation}
\begin{aligned}
 X(t)&=[\boldsymbol{u}^{(\alpha_1)}(\boldsymbol{x},t),\,
        \boldsymbol{u}^{(\alpha_2)}(\boldsymbol{x},t),\,\cdots,\,
        \boldsymbol{u}^{(\alpha_{j_{\text{max}}})}(\boldsymbol{x},t)]
        \in\mathbb{R}^{N\times j_{\text{max}}},\\
 Y(t)&=[\boldsymbol{u}^{(\alpha_1)}(\boldsymbol{x},t+\Delta T),\,
        \boldsymbol{u}^{(\alpha_2)}(\boldsymbol{x},t+\Delta T),\,\cdots,\,
        \boldsymbol{u}^{(\alpha_{j_{\text{max}}})}(\boldsymbol{x},t+\Delta T)]
        \in\mathbb{R}^{N\times j_{\text{max}}}.
\end{aligned}
\label{matrixX_general}
\end{equation}
where $\boldsymbol{u}^{(\alpha_j)}(\boldsymbol{x},t)$ is the realization with phase $\alpha_j$. The associated tDMDpc modes are denoted by $\boldsymbol{\varphi}_k(\boldsymbol{x},t)$, to distinguish them from the stationary POD modes $\boldsymbol{\phi}_k$. The state consists of both velocity components, weighted by the cell areas; the phase-ensemble mean is not subtracted. In the SVD $W^{1/2}X=U_rS_rV_r^*$, the reduced operator is $\widetilde A=U_r^*W^{1/2}YV_rS_r^{-1}$. We use the projected-DMD modes\cite{Tu2014} $\boldsymbol{\varphi}_k=W^{-1/2}U_rz_k$, where $z_k$ is an eigenvector of $\widetilde A$. The growth rate $\sigma_k$ and frequency $f_k$ are obtained from $\lambda_k^c(t)=\sigma_k(t)+2\pi i f_k(t)=\log\mu_k(t)/\Delta T$, where $\mu_k(t)$ is an eigenvalue of $\widetilde A(t)$\cite{Rowley2009}. Each mode is normalized to $\|\boldsymbol{\varphi}_k\|=1$. Its arbitrary complex phase does not affect the modal contribution in Eq.~\eqref{definecoherent}.

At each time, the reconstructed velocity field $\boldsymbol{u}_{\mathrm{rec}}^{(\alpha)}$ for phase $\alpha$ is obtained from the retained tDMDpc modes as
\begin{equation}
\begin{aligned}
 \boldsymbol{u}^{(\alpha)}(\boldsymbol{x},t)
 &\approx\boldsymbol{u}_{\mathrm{rec}}^{(\alpha)}(\boldsymbol{x},t)=\sum_{k\in K}a_k^{(\alpha)}(t)
   \boldsymbol{\varphi}_k(\boldsymbol{x},t).
\end{aligned}
\label{DMDrepresent}
\end{equation}
where $K$ contains the retained modes and $|K|=r$, with $r$ denoting the SVD rank. In the retained sets used here, $K$ contains the zero-frequency mode ($k=0$) and $(r-1)/2$ complex-conjugate pairs; $k=\pm1$ labels the fundamental pair. The coefficient vector is obtained independently for each realization by spatially weighted least squares,
\begin{equation}
\begin{aligned}
 \boldsymbol{a}^{(\alpha)}(t)
 &=\left[W^{1/2}\Phi(t)\right]^\dagger
   W^{1/2}\boldsymbol{u}^{(\alpha)}(t),\\
 \Phi(t)&=[\boldsymbol{\varphi}_k(t)]_{k\in K}.
\end{aligned}
 \label{eq:coefficients}
\end{equation}
where $\dagger$ denotes the Moore--Penrose pseudoinverse. Because the modes are not orthogonal, the coefficients are fitted simultaneously rather than by projecting onto each mode separately.

If the realizations were phase-shifted copies of the same periodic orbit, the coefficients could take the form $a_k^{(\alpha)}(t)=b_k(t)e^{ik\alpha}$, with the sign determined by the phase convention. In the gust encounter, each realization meets the same gust at a different shedding phase, so the magnitude and phase of its coefficients evolve differently. Here $\alpha$ labels the realization, while the argument of $a_k^{(\alpha)}$ gives its instantaneous coefficient phase. One common operator is estimated from the ensemble, and the phase-specific coefficients describe the differences among the realizations.
We define the modal contribution by
\begin{equation}
 \hat{\boldsymbol{u}}_k^{(\alpha)}(\boldsymbol{x},t)
 =a_k^{(\alpha)}(t)\boldsymbol{\varphi}_k(\boldsymbol{x},t).
 \label{definecoherent}
\end{equation}
Because the velocity is real, a positive-frequency contribution and its complex conjugate combine into the real pair contribution $2\operatorname{Re}\hat{\boldsymbol{u}}_k^{(\alpha)}$. The contribution in Eq.~\eqref{definecoherent} is independent of the mode normalization, whereas the coefficient alone is not. For comparisons across phases, we show coefficient magnitudes divided by their phase-ensemble mean.

\subsection{Modal lift contributions}\label{sec:modal_lift}
To connect the evolving flow structures to the aerodynamic forces, we evaluate the lift and its modal contributions. For each realization, the pressure and viscous stresses are integrated over the cylinder surface to give the vertical force $F_y$, and the lift coefficient is
\begin{equation}
 C_L=\frac{F_y}{\tfrac12\rho U_\infty^2D}.
 \label{eq:lift_coefficient}
\end{equation}

To obtain the modal lift contributions, we apply tDMDpc to velocity and pressure using the state
\begin{equation*}
 \left(\frac{u_x}{U_\infty},\frac{u_y}{U_\infty},
 \frac{p}{\rho U_\infty^2}\right)^\top.
\end{equation*}
We apply the same area-weighted inner product to all three nondimensional components. We retain $r=11$ modes from the $32$-phase ensemble, normalized to unit norm. Applying Eq.~\eqref{eq:coefficients} to this state gives one set of phase-specific coefficients for both velocity and pressure. The zero-frequency and fundamental modes are identified by comparing their velocity components with the corresponding modes obtained from velocity fields alone. The fundamental frequency and growth rate have time histories similar to those obtained from velocity alone. For each phase, we reconstruct the zero-frequency contribution and the fundamental conjugate pair. We integrate the pressure and viscous stresses of these reconstructed fields over the cylinder surface and normalize the forces using Eq.~\eqref{eq:lift_coefficient}. The resulting lift contributions are denoted by $C_{L,0}^{(\alpha)}$ and $C_{L,1}^{(\alpha)}$, respectively. Summing the lift contributions of all retained modes gives the reconstructed lift $C_{L,\mathrm{rec}}^{(\alpha)}$.

\section{Numerical setup}\label{sec:numerical}
\subsection{Numerical methods}\label{subsec:numerical}
\begin{figure}[H]
\centering
\includegraphics[width=8.5cm,keepaspectratio]{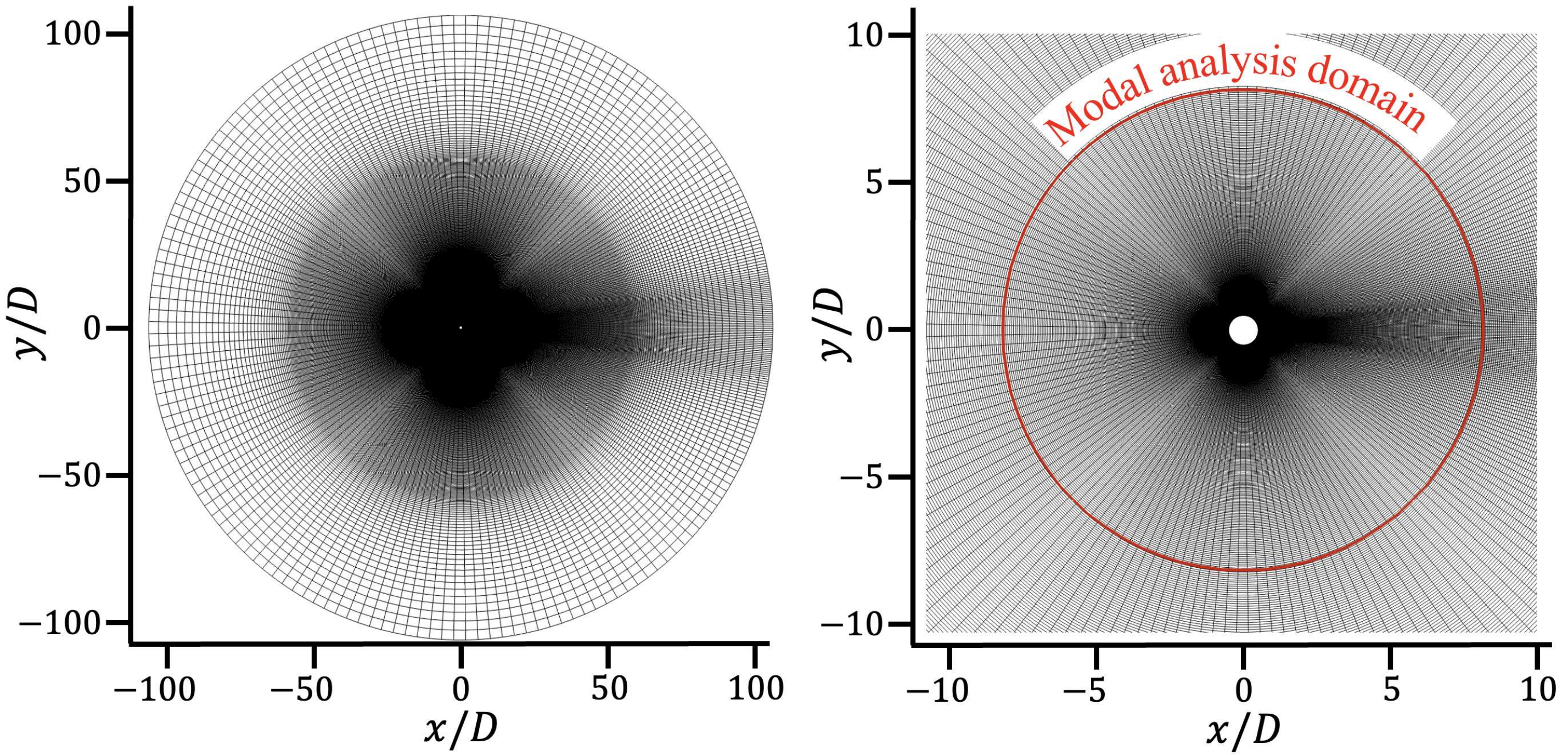}

\captionsetup{justification=raggedright,singlelinecheck=false}
\captionsetup{width=8.5cm,skip=2pt}
\caption{Computational grid around a circular cylinder and modal analysis domain. (a) Overall view. (b) Close-up view.}
\label{grid}

\end{figure}

The flow around a circular cylinder was computed by solving the incompressible Navier--Stokes equations,
\begin{equation}
\nabla\cdot\boldsymbol{u}=0,
   \label{eqcont}	
\end{equation}
\begin{equation}
\frac{\partial{\boldsymbol{u}}}{\partial{t}}=-(\boldsymbol{u}\cdot\nabla)\boldsymbol{u}-\frac{1}{\rho}\nabla{p}+\frac{1}{Re}{\nabla^2}\boldsymbol{u},
   \label{eqnavi}
\end{equation}
where $\boldsymbol{u}=(u_x,u_y)^\top$ represents the velocity vector, ${p}$ is the pressure, and $\rho$ is the fluid density. $Re$ is the Reynolds number defined as
\begin{equation}
Re  \stackrel{\mathrm{def}}{=}  \frac{U_{\infty}D}{\nu},
   \label{eqRe}
\end{equation}
where $U_{\infty}$ denotes the freestream velocity, $\nu$ is the kinematic viscosity, and $D$ is the cylinder diameter. 
The numerical procedures, including the boundary conditions, are described in our previous work\cite{mypaper2,mypaper3}. Figure~\ref{grid} shows the computational grid, with $240$ points in the circumferential direction and $590$ in the radial direction. The modal analysis uses both velocity components within the region marked by the red circle in Fig.~\ref{grid}(b). Snapshots are sampled at intervals of $0.1D/U_\infty$, which is also the time separation used in tDMDpc. The $64$ phases are uniformly spaced over one shedding cycle, and the $32$-phase ensemble uses every other phase.

\subsection{Vortex gust model}\label{sec:gust_model}
The gust is modeled as an isolated Taylor vortex specified by a Gaussian stream function and superimposed upstream of the cylinder\cite{MartinezMurielFlores2020}. This model has a divergence-free velocity field and zero net circulation over the unbounded plane. The stream function is
\begin{equation}
\begin{aligned}
 \psi(\ell_g)&=\psi_0\exp\!\left(-\frac{\ell_g^2}{2R^2}\right),\qquad \ell_g=\lvert\boldsymbol{x}-\boldsymbol{x}_0\rvert.
\end{aligned}
  \label{eq:gust_psi}
\end{equation}
where $\boldsymbol{x}_0=(x_0 ,y_0)$ is the vortex center and $R$ is the gust radius.
The corresponding velocity field, $\boldsymbol{u}_g=(\partial\psi/\partial y,\,-\partial\psi/\partial x)$,
is purely azimuthal,
\begin{equation}
  u_\theta(\ell_g) = u_{\theta,\max}\,\frac{\ell_g}{R}\,
  \exp\!\left(\frac{1}{2}-\frac{\ell_g^{2}}{2R^{2}}\right),
  \label{eq:gust_utheta}
\end{equation}
and attains its peak value $u_{\theta,\max}=\psi_0/(R\sqrt{e})$ at $\ell_g=R$.
In Cartesian components,
\begin{align}
 u_{g,x}&=-u_{\theta,\max}\frac{y-y_0}{R}\,\times\exp\!\left(\frac12-
 \frac{(x-x_0)^2+(y-y_0)^2}{2R^2}\right),
 \label{eq:gust_ux}\\
 u_{g,y}&=+u_{\theta,\max}\frac{x-x_0}{R}\,\times\exp\!\left(\frac12-
 \frac{(x-x_0)^2+(y-y_0)^2}{2R^2}\right).
 \label{eq:gust_uy}
\end{align}
For $\psi_0>0$, the prescribed vortex rotates counterclockwise. Its spanwise vorticity is
\begin{equation}
 \omega_{g,z}(\ell_g)=\left(\frac{2}{R^2}-\frac{\ell_g^2}{R^4}\right)\psi_0\exp\!\left(-\frac{\ell_g^2}{2R^2}\right).
\end{equation}
The positive-vorticity core is surrounded by a negative-vorticity annulus outside $\ell_g=\sqrt{2}R$. Both regions belong to the gust and can deform the wake.

The prescribed gust parameters are $(G,R,x_0,y_0)=(1,0.5D,-10D,1D)$, so the introduced vortex has $u_{\theta,\max}=U_\infty$. The swirl velocity peaks on a circle of diameter $2R=D$. The vortex is introduced at $t=t_0$ after the wake has reached its periodic state; $\alpha$ denotes the shedding phase at introduction. We use $\tau=t-t_0$ for the elapsed time in units of $D/U_\infty$, with $\tau=10$ as the nominal arrival. Figure labels of the form $t=t_0+\tau$ refer to the same elapsed time.

\section{Results and discussion}\label{sec:result}
\subsection{Stationary POD and the phase-dependent response}\label{sec:flowresult}
The undisturbed periodic wake at $Re=100$ provides the reference for identifying departures during the gust encounter. Its Strouhal number is approximately $0.164$ and its lift-coefficient amplitude is about $0.32$, consistent with previous results\cite{Mittal2005}; one shedding period is about $6.1D/U_\infty$. We first characterize this reference with stationary POD and then compare the trajectories at different phases during the encounter.

\begin{figure}[H]
\centering
\includegraphics[width=8.9cm,trim=0bp 435bp 0bp 0bp,clip]{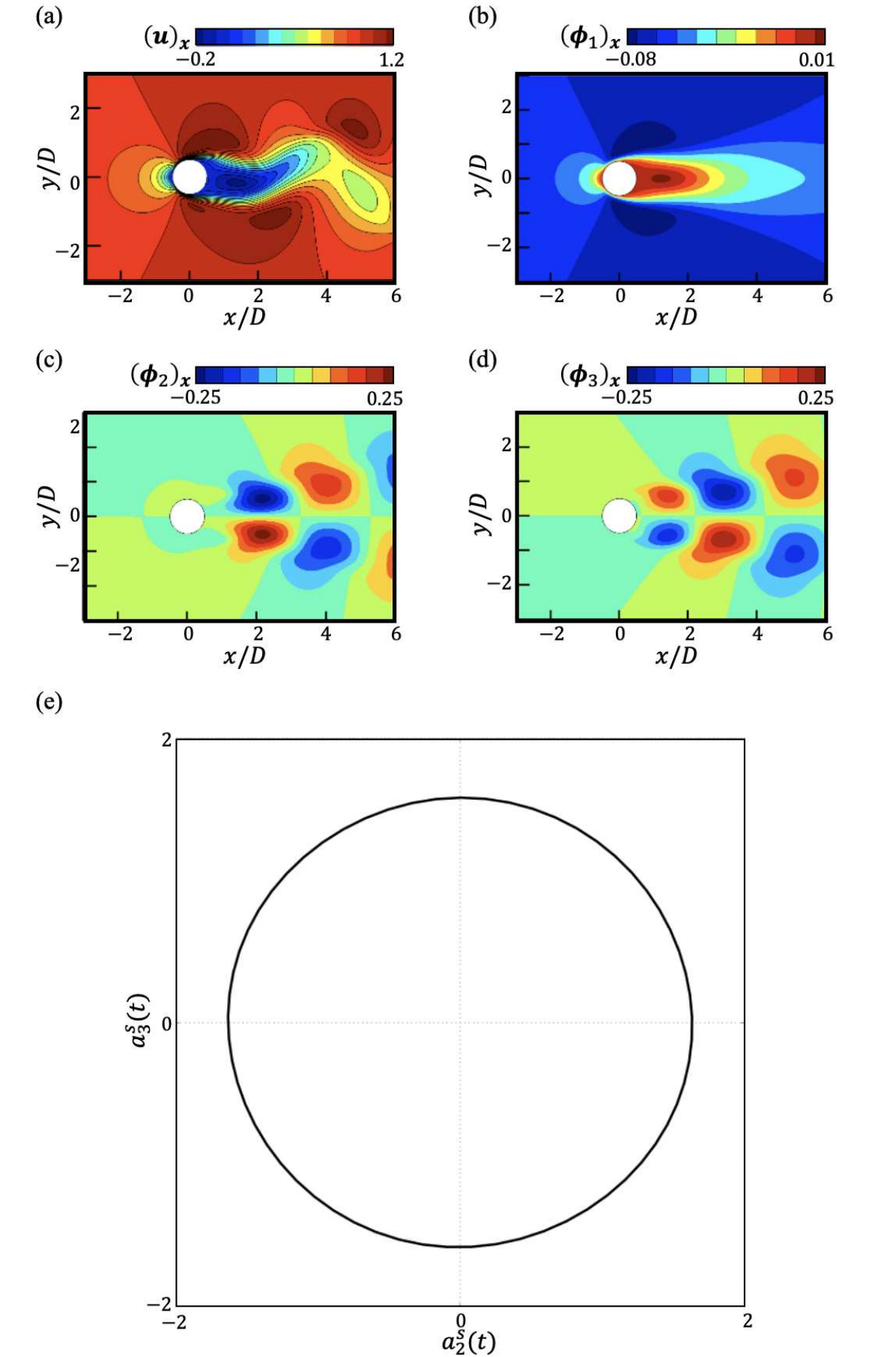}\hfill
\includegraphics[width=7.3cm,trim=15bp 0bp 70bp 435bp,clip]{fig/fig1.pdf}

\captionsetup{justification=raggedright,singlelinecheck=false}
\captionsetup{width=\linewidth,skip=2pt}
\caption{Periodic cylinder wake before the introduction of the vortex gust. (a) Instantaneous streamwise velocity field. (b–d) Streamwise velocity component of the first three stationary POD modes, $\boldsymbol{\phi}_k(\boldsymbol{x})$ ($k=1, 2, 3$). (e) Phase portrait of the corresponding POD coefficients $a^s_k(t)$ computed from the periodic cylinder wake.}
\label{fig1}
\end{figure}

\begin{figure}[H]
\centering
\includegraphics[width=12.5cm,keepaspectratio]{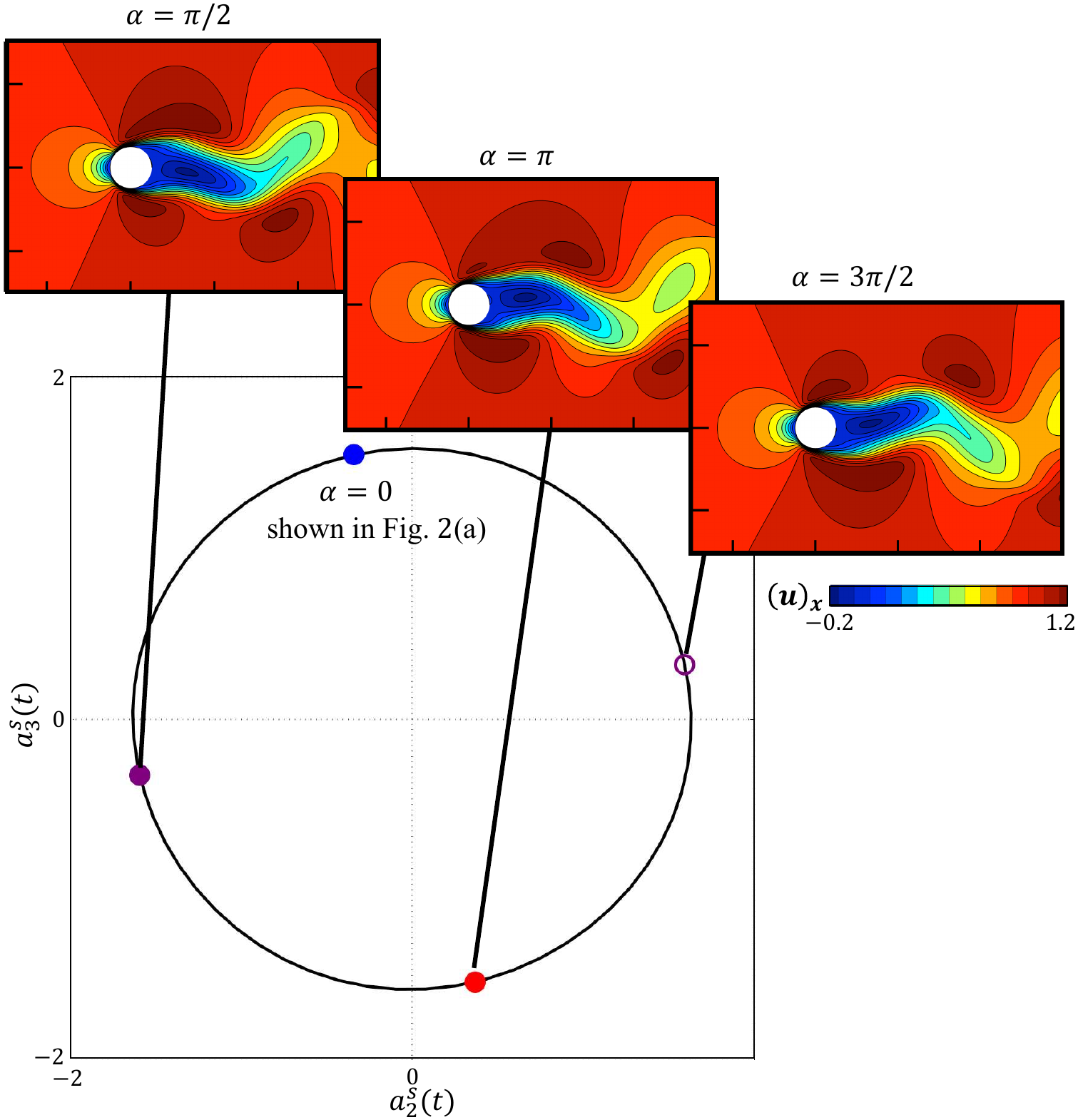}

\captionsetup{justification=raggedright,singlelinecheck=false}
\captionsetup{width=12.5cm,skip=2pt}
\caption{Instantaneous streamwise velocity fields of the periodic cylinder wake at $\alpha=\pi/2, \pi$, and $3\pi/2$ before the introduction of the vortex gust.}
\label{fig2}
\end{figure}

Figure \ref{fig1}(a) shows an instantaneous streamwise velocity field before the gust is introduced. POD was applied to the periodic state, and the resulting POD modes $\boldsymbol{\phi}_k(\boldsymbol{x})$ are shown in Fig. \ref{fig1}(b–d). The first POD mode, $\boldsymbol{\phi}_1(\boldsymbol{x})$, captures the mean flow, whereas $\boldsymbol{\phi}_2(\boldsymbol{x})$ and $\boldsymbol{\phi}_3(\boldsymbol{x})$ form a pair that represents the vortex shedding. Figure \ref{fig1}(e) shows the trajectory of their coefficients, which traces a limit cycle\cite{Noack2003}. These stationary POD modes provide a common reference for comparing the disturbed wakes.
To examine different phases, we choose the state shown in Fig.~\ref{fig1}(a) as the phase reference and assign it $\alpha=0$. The periodic states at the other phases are obtained using the procedure described in our previous work\cite{mypaper_6}. Figure~\ref{fig2} shows representative initial states. The near-wake vortices occupy different positions relative to the cylinder, so the same gust meets a different local vortex configuration in each case.

\begin{figure}[H]
\centering
\includegraphics[width=\linewidth,keepaspectratio]{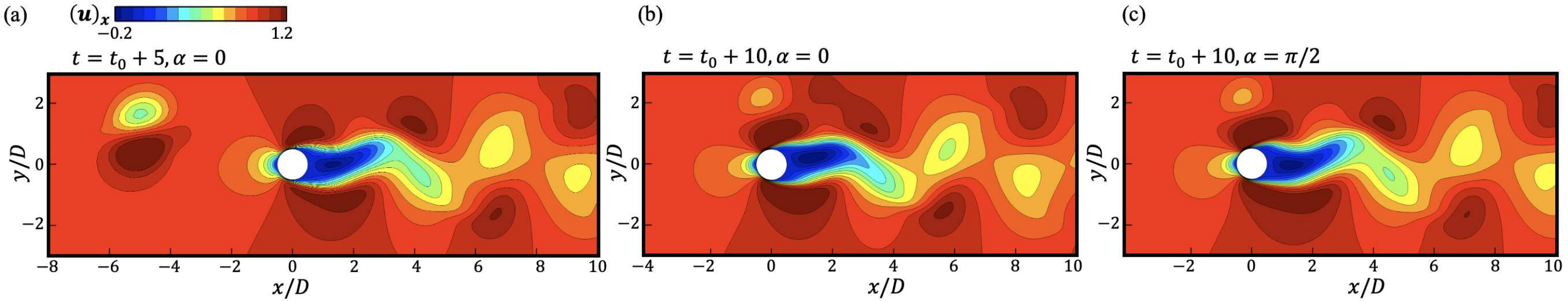}

\captionsetup{justification=raggedright,singlelinecheck=false}
\captionsetup{width=\linewidth,skip=2pt}
\caption{Instantaneous streamwise velocity fields showing the interaction between the vortex gust and the cylinder wake. (a,b) $\alpha=0$ at $t=t_0+5$ and $t=t_0+10$. (c) $\alpha=\pi/2$ at $t=t_0+10$.}
\label{fig3}

\end{figure}

Figure~\ref{fig3} shows the approach and encounter: the gust lies upstream of the cylinder at $\tau=5$ and interacts with the near wake at $\tau=10$. At the same elapsed time, the near-wake vortices occupy different positions for $\alpha=0$ and $\pi/2$.

\begin{figure}[H]
\centering
\includegraphics[width=15.5cm,keepaspectratio]{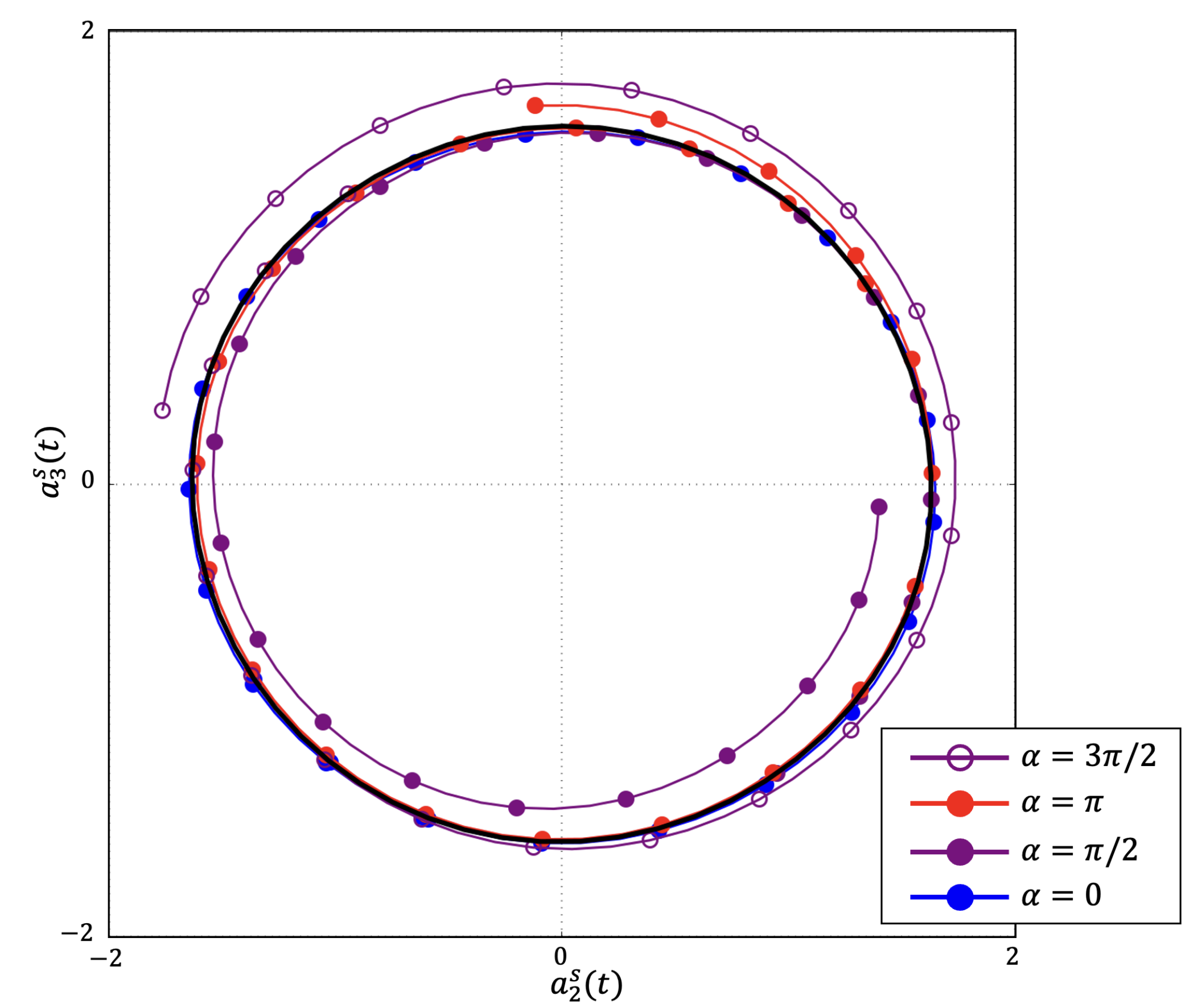}

\captionsetup{justification=raggedright,singlelinecheck=false}
\captionsetup{width=15.5cm,skip=2pt}
\caption{Phase portraits of the POD coefficients associated with the stationary POD modes,  $\boldsymbol{\phi}_2(\boldsymbol{x})$ and $\boldsymbol{\phi}_3(\boldsymbol{x})$, for $\alpha=0, \pi/2, \pi$, and $3\pi/2$, over $8\leq\tau\leq14.9$.}
\label{fig4}

\end{figure}

Figure~\ref{fig4} compares the disturbed trajectories in the stationary POD plane for four representative phases over $8\leq\tau\leq14.9$. The trajectories move outward or inward depending on the phase, most clearly after arrival. During the interaction, the gust also changes the spatial distribution of the Kármán vortices. We therefore apply tDMDpc to examine the evolving structures associated with these different amplitude responses.

\subsection{tDMDpc-based analysis}
We apply tDMDpc at each elapsed time using SVD rank $r=11$ and the $32$-phase ensemble unless otherwise noted; the $64$-phase ensemble is used to check the sensitivity to ensemble size in Fig.~\ref{fig5}.
We first evaluate the reconstruction errors of $\boldsymbol{u}_{\mathrm{rec}}^{(\alpha)}$ in Eq.~\eqref{DMDrepresent} as
\begin{equation}
\begin{aligned}
 \epsilon_u^{(\alpha)}(t)
 &=\frac{\|\boldsymbol{u}^{(\alpha)}-
       \boldsymbol{u}_{\mathrm{rec}}^{(\alpha)}\|}
       {\|\boldsymbol{u}^{(\alpha)}\|},\\
 \epsilon_{\mathrm{fluc}}^{(\alpha)}(t)
 &=\frac{\|\boldsymbol{u}^{(\alpha)}-
       \boldsymbol{u}_{\mathrm{rec}}^{(\alpha)}\|}
       {\|\boldsymbol{u}^{(\alpha)}-\overline{\boldsymbol{u}}\|}.
\end{aligned}
 \label{eq:rec_errors}
\end{equation}
where the overbar denotes the phase-ensemble mean.

\begin{figure}[H]
\centering
\includegraphics[width=9.5cm]{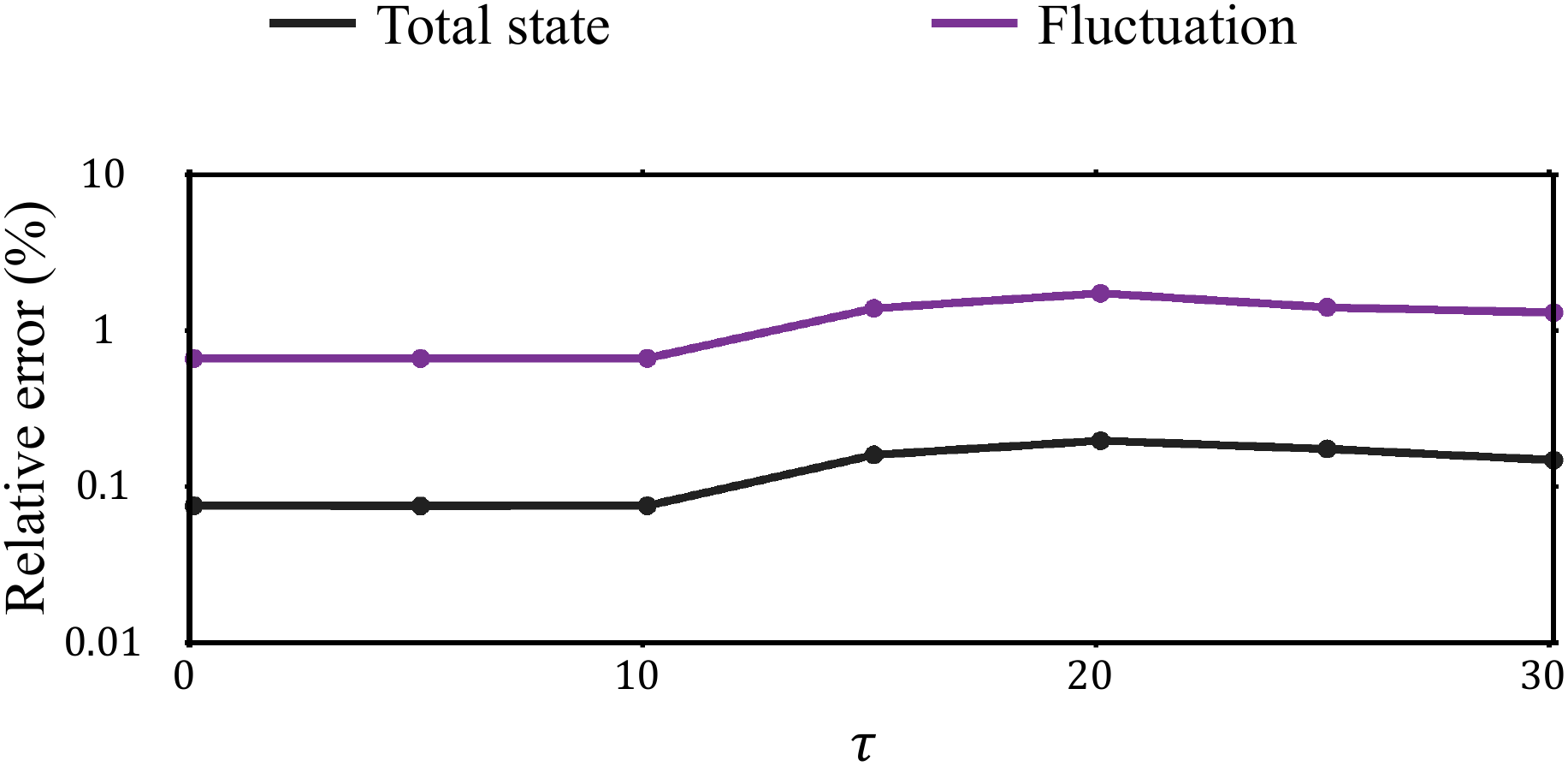}
\captionsetup{width=9.5cm,skip=2pt,justification=raggedright,singlelinecheck=false}
\caption{Maximum relative reconstruction errors across the $32$ represented phases for $r=11$, normalized by the total state (black) and the phase-dependent fluctuation (purple).}
\label{fig:reconstruction_error}
\end{figure}

\begin{figure}[H]
\centering
\includegraphics[width=15cm,keepaspectratio]{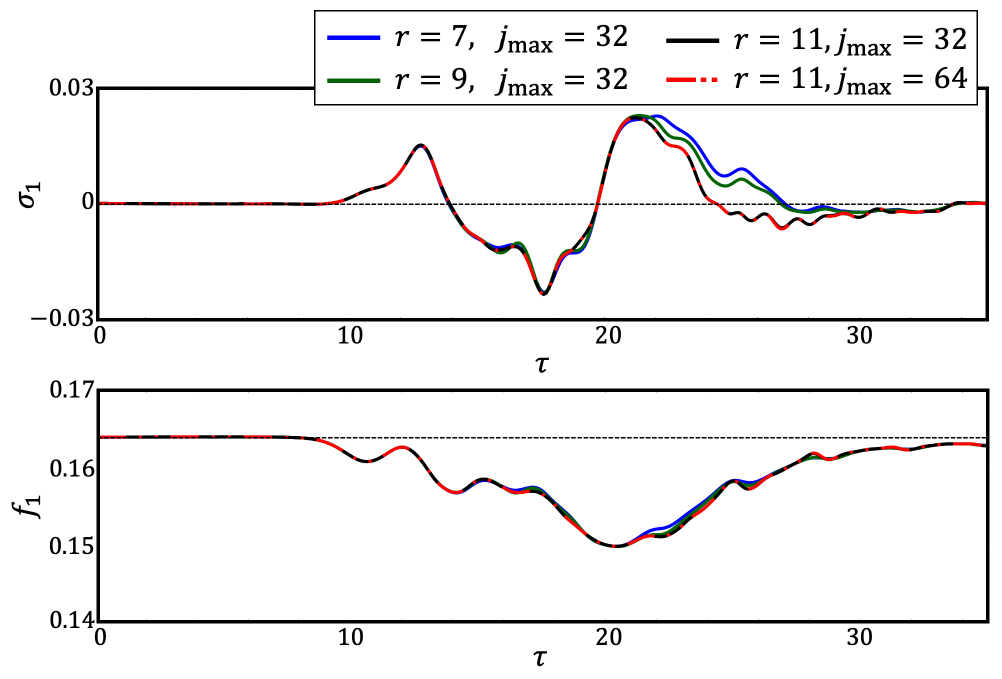}

\captionsetup{justification=raggedright,singlelinecheck=false}
\captionsetup{width=\linewidth,skip=2pt}
\caption{Time variation of the frequency and growth rate of the fundamental tDMDpc mode for different SVD ranks ($r=7,9,11$) and phase-ensemble sizes ($j_\text{max}=32,64$).}
\label{fig5}

\end{figure}

Normalizing by the fluctuation prevents the large mean velocity from masking the error. Because the retained projected-DMD modes span the rank-$r$ SVD subspace, these residuals measure its truncation error rather than the accuracy of the estimated operator. At the sampled times before, during, and after the encounter, the maximum error over the $32$ phases stays below $0.2\%$ of the total state and $2\%$ of the fluctuation [Fig.~\ref{fig:reconstruction_error}], indicating that the retained modes represent the transient velocity fields across these stages.

Figure~\ref{fig5} shows the time histories of the fundamental-mode frequency $f_1$ and growth rate $\sigma_1$ obtained for different SVD ranks and phase-ensemble sizes. The frequency and growth rate agree across the tested ranks and ensemble sizes up to $\tau\simeq20$, about $1.6$ shedding periods after arrival. Beyond this interval, the estimate depends on the rank and the ensemble size. The temporary decrease in $f_1$, from approximately $0.164$ to $0.150$ near $\tau=20$, indicates a slowing of the estimated fundamental oscillation; the frequency returns close to its pre-encounter value around $\tau=30$. Positive and negative values of $\sigma_1$ indicate instantaneous growth and decay under the estimated operator. Because the modes are time dependent and nonorthogonal, $\sigma_1$ does not in general equal the growth rate of $|a_1|$ for an individual phase.

\begin{figure}[H]
\centering
\includegraphics[width=\linewidth,keepaspectratio]{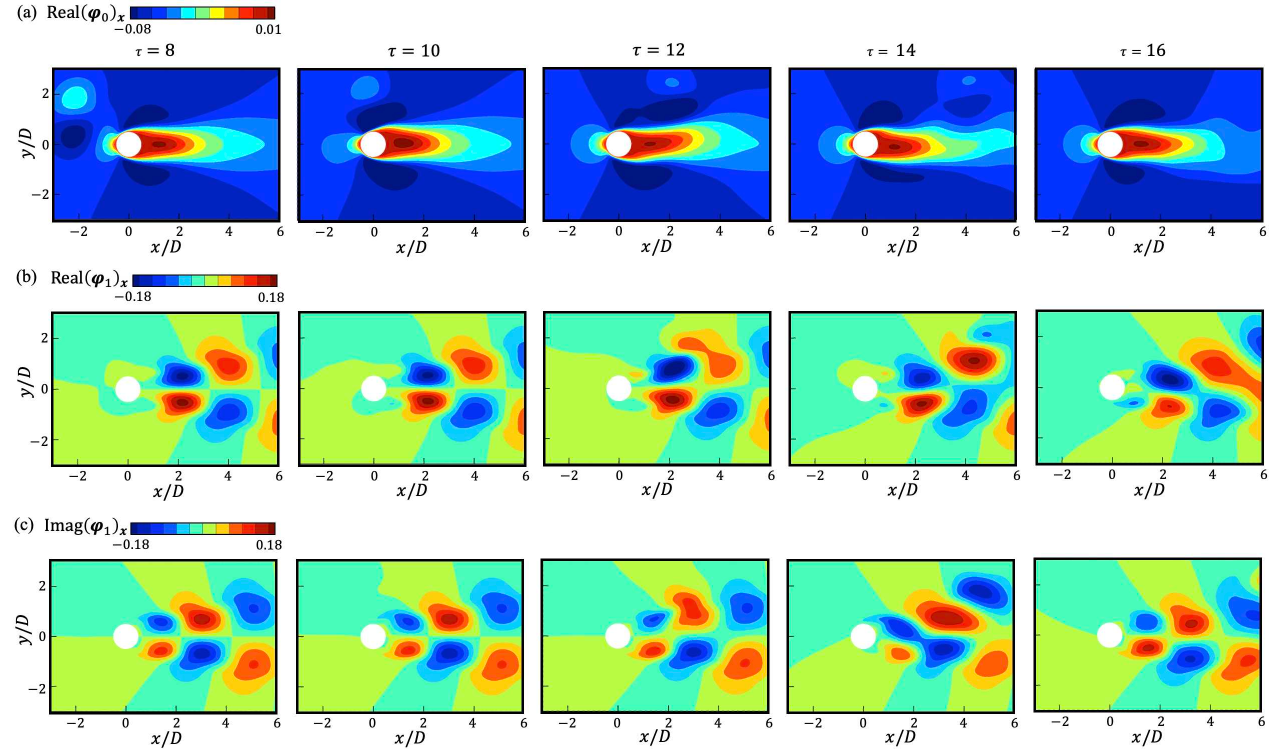}

\captionsetup{justification=raggedright,singlelinecheck=false}
\captionsetup{width=\linewidth,skip=2pt}
\caption{Temporal evolution of the streamwise components of the tDMDpc velocity modes during the vortex-gust interaction: (a) the real part of the zero-frequency mode, $\operatorname{Re}(\boldsymbol{\varphi}_0)_x$, and (b) the real and (c) the imaginary parts of the fundamental mode, $\operatorname{Re}(\boldsymbol{\varphi}_1)_x$ and $\operatorname{Im}(\boldsymbol{\varphi}_1)_x$. Each velocity mode has unit norm. Contour levels are fixed within each row.}
\label{fig6}

\end{figure}

\begin{figure}[H]
\centering
\includegraphics[width=\linewidth,keepaspectratio]{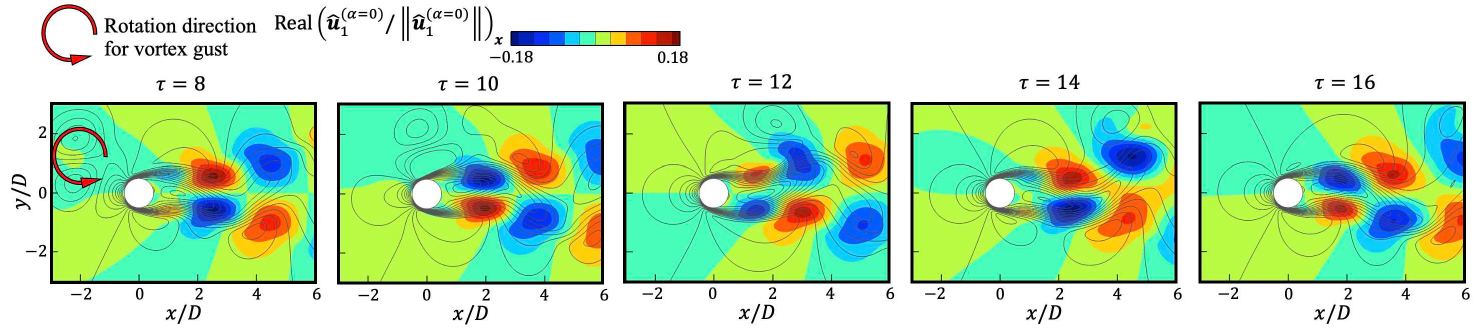}

\captionsetup{justification=raggedright,singlelinecheck=false}
\captionsetup{width=\linewidth,skip=2pt}
\caption{Time variation of the real part of the streamwise component of the normalized fundamental-mode contribution for $\alpha=0$. Filled contours show $\operatorname{Re}[\hat{\boldsymbol{u}}_1^{(0)}/\|\hat{\boldsymbol{u}}_1^{(0)}\|]_x$; line contours show the instantaneous streamwise velocity.}
\label{fig7}

\end{figure}

Figure~\ref{fig6} shows the temporal evolution of the zero-frequency mode and the fundamental mode associated with vortex shedding. Before arrival, the approaching gust appears mainly in the zero-frequency mode [Fig.~\ref{fig6}(a)]. During the interaction, gust-related structures also appear in the fundamental mode [Fig.~\ref{fig6}(b,c)]. These evolving modes capture both the downstream convection of the gust and the changes near the cylinder.
To relate the phase-dependent amplitude responses to the wake structures, we next examine the fundamental-mode contributions at different phases. For $\alpha=0$, Fig.~\ref{fig7} overlays the normalized fundamental-mode contribution with contours of the instantaneous streamwise velocity. Before arrival, the main positive and negative regions of the contribution are concentrated in the shedding wake, while the approaching gust lies upstream of the cylinder. At $\tau=12$, the upper negative lobe of the fundamental contribution extends upward toward the gust-related velocity contours. At $\tau=14$--$16$, the distorted region lies farther downstream, while the alternating positive and negative pattern remains near the cylinder. The contribution thus captures both the gust-affected region as it moves downstream and the shedding that continues near the cylinder.

\begin{figure}[H]
\centering
\includegraphics[width=14.5cm,keepaspectratio]{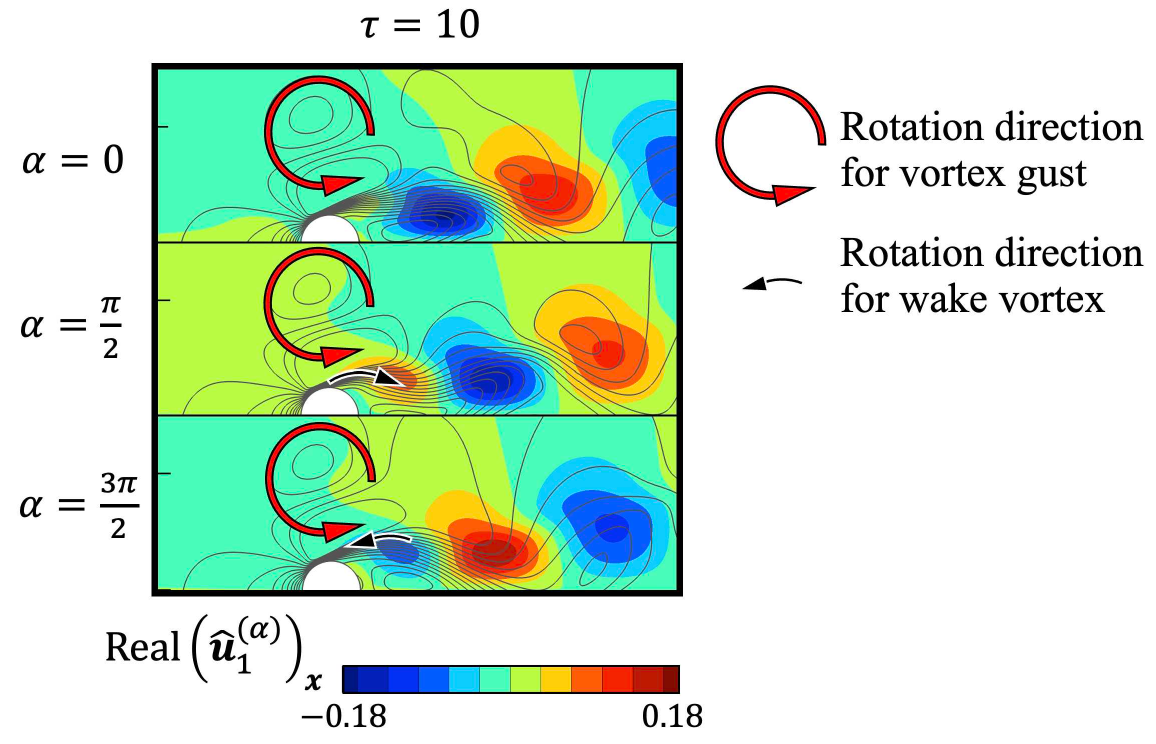}

\captionsetup{justification=raggedright,singlelinecheck=false}
\captionsetup{width=14.5cm,skip=2pt}
\caption{Real part of the streamwise component of the fundamental-mode contributions at $\tau=10$ for $\alpha=0,\pi/2,3\pi/2$. Filled contours show $\operatorname{Re}(\hat{\boldsymbol{u}}^{(\alpha)}_1)_x$; line contours show the instantaneous streamwise velocity. The red arrow indicates gust rotation; black arrows indicate rotation in the fundamental-mode contributions.}
\label{fig8}
\end{figure}

Figure~\ref{fig8} compares the fundamental-mode contributions at $\tau=10$ for different phases. The red arrow indicates the rotation of the vortex gust, and the black arrows indicate rotation in the fundamental-mode contributions. The counter-rotating and co-rotating configurations at $\alpha=\pi/2$ and $3\pi/2$ are consistent with the attenuation and amplification observed in the POD amplitude, respectively [Fig.~\ref{fig4}].

\begin{figure}[H]
\centering
\includegraphics[width=12cm]{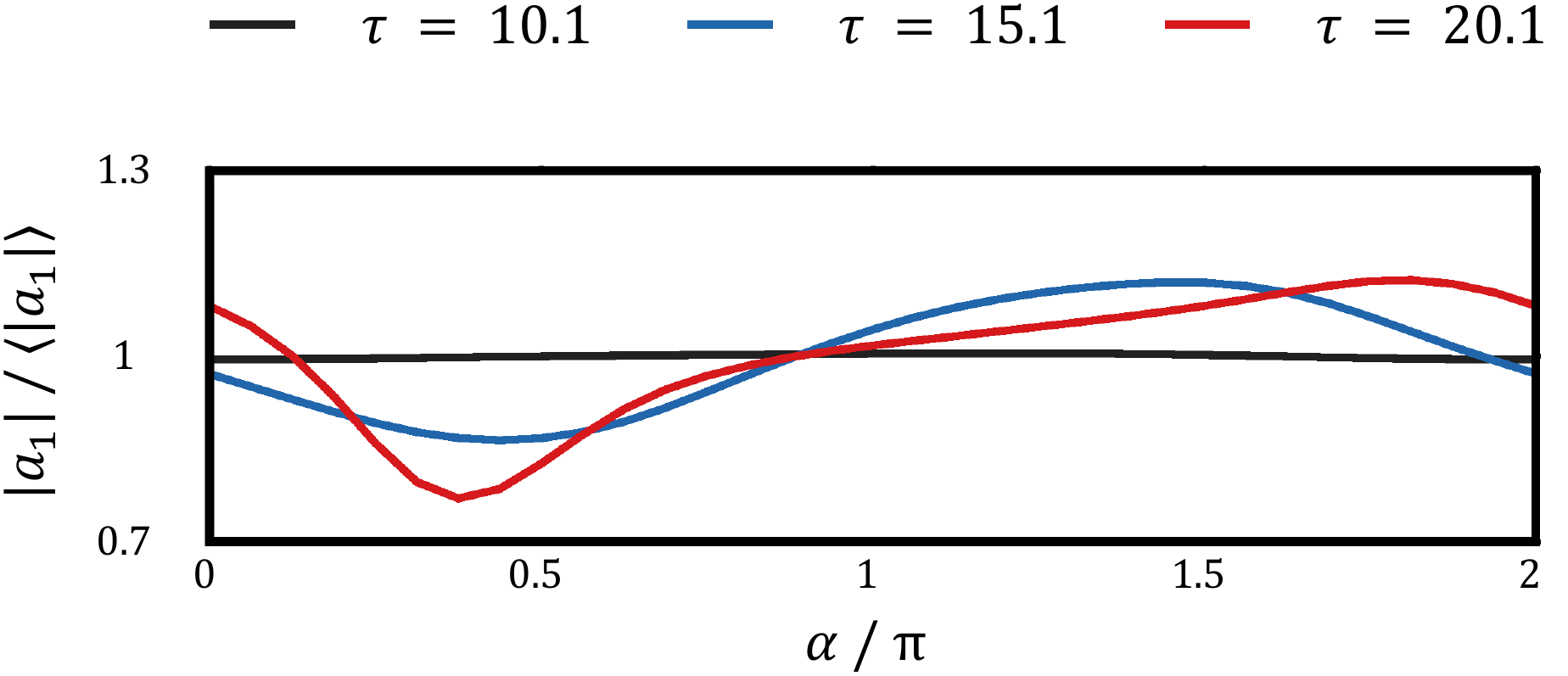}
\captionsetup{width=\linewidth,skip=2pt,justification=raggedright,singlelinecheck=false}
\caption{Fundamental-mode coefficient magnitude divided by its phase-ensemble mean at $\tau=10.1$ (black), $15.1$ (blue), and $20.1$ (red), using $r=11$ and $32$ phases.}
\label{fig:modal_validation}
\end{figure}

The phase-specific coefficients quantify the strength of a mode in each realization. Figure~\ref{fig:modal_validation} compares the normalized fundamental-mode coefficient magnitude across phases. It is nearly uniform at $\tau=10.1$, and clear phase variation appears at $\tau=15.1$ and $20.1$. At these later times, the coefficient magnitude lies below the phase-ensemble mean for $\pi/2$ and above it for $3\pi/2$. These are the phases whose POD trajectories move inward and outward in Fig.~\ref{fig4}. The decrease around $\pi/2$ is more pronounced at $\tau=20.1$ than at $15.1$. The same gust therefore weakens the fundamental shedding contribution near $\pi/2$ and strengthens it near $3\pi/2$.


Figure~\ref{fig:modal_validation} compares the phases at fixed times. To follow each phase in time, we evaluate the vorticity amplitude of the fundamental component in the near wake. The local amplitude envelope is $2|(\nabla\times\hat{\boldsymbol{u}}_1^{(\alpha)})_z|$, obtained from the complex modal contribution in Eq.~\eqref{definecoherent}. We compute its area-weighted spatial root-mean-square over $0.5<x/D<4$ and $|y|/D<2$, and normalize each history by its value at $\tau=5.1$, before the encounter.

\begin{figure}[H]
\centering
\includegraphics[width=14cm]{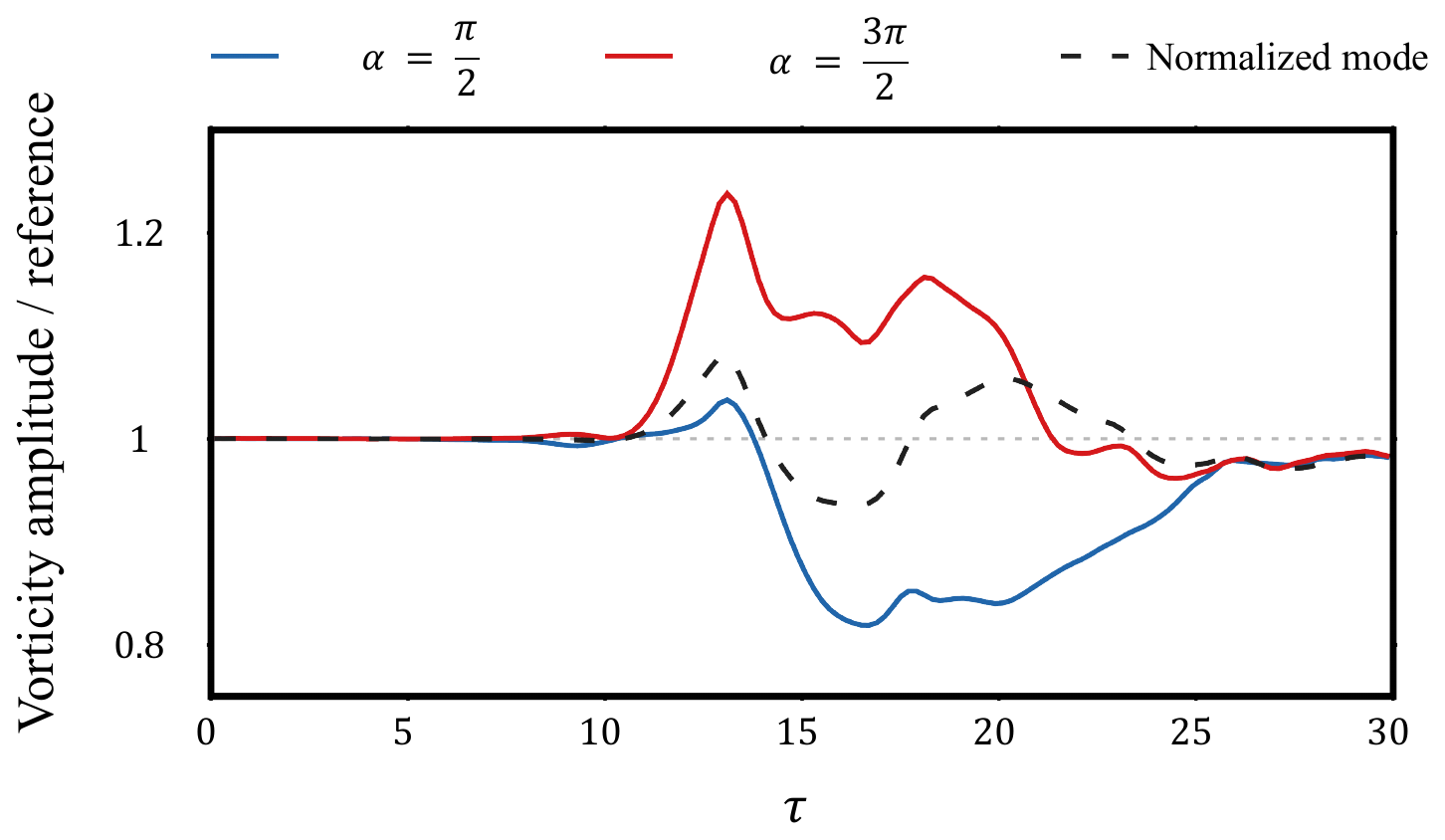}
\captionsetup{width=\linewidth,skip=2pt,justification=raggedright,singlelinecheck=false}
\caption{Fundamental-mode spanwise vorticity amplitude in the near wake, calculated from both velocity components. Solid curves show the spatial root-mean-square of the local amplitude envelope of the conjugate pair for $\alpha=\pi/2$ (blue) and $3\pi/2$ (red). The black dashed curve shows the spatial root-mean-square of $|(\nabla\times\boldsymbol{\varphi}_1)_z|$ for the unit-norm complex fundamental velocity mode. All curves use area-weighted spatial averages over $0.5<x/D<4$ and $|y|/D<2$ and are normalized by their own values at $\tau=5.1$. The horizontal dotted line marks this pre-encounter reference.}
\label{fig:vorticity_amplitude}
\end{figure}

Figure~\ref{fig:vorticity_amplitude} shows that both amplitudes remain close to their pre-encounter values until arrival and then separate. Over approximately $14\lesssim\tau\lesssim21$, the amplitude decreases by up to about $18\%$ for $\alpha=\pi/2$ and increases by up to about $16\%$ for $3\pi/2$ relative to their respective pre-encounter values. This change shows that the fundamental shedding component is attenuated at $\pi/2$ and amplified at $3\pi/2$, consistent with the stationary POD trajectories and the phase-specific coefficients. The sharper peak near $\tau=13$ coincides with a peak in the dashed curve and partly reflects a change in the spatial distribution of the mode. At later times, the two histories approach the reference level and the difference between them decreases.

\subsection{Modal contributions to the transient lift}

\begin{figure}[H]
\centering
\includegraphics[width=13cm]{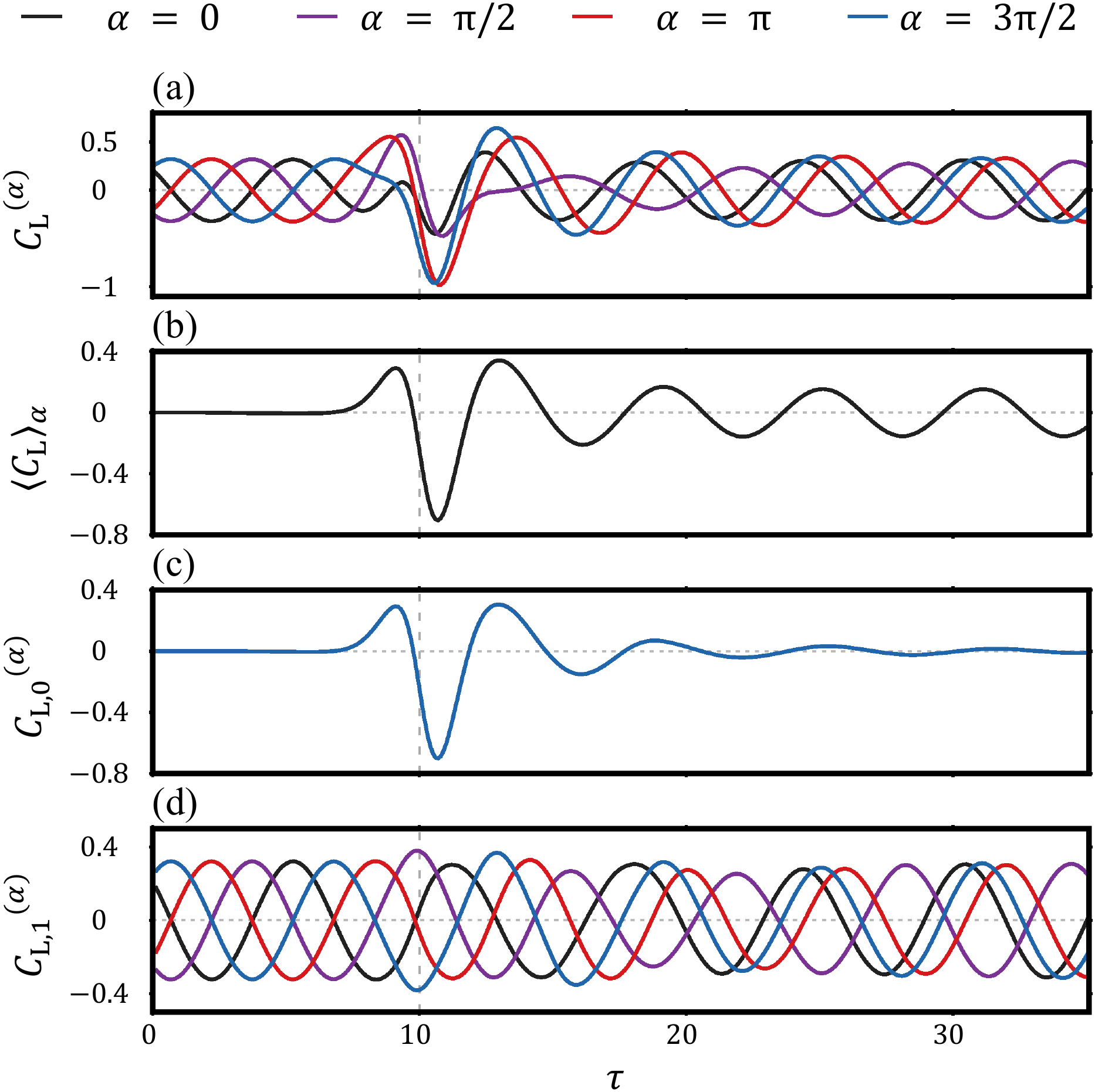}
\captionsetup{width=\linewidth,skip=2pt,justification=raggedright,singlelinecheck=false}
\caption{Transient lift and its modal decomposition. (a) Total lift coefficient for $\alpha=0,\pi/2,\pi,3\pi/2$. (b) Phase-ensemble mean over all $64$ phases at each elapsed time. Surface-integrated contributions of (c) the zero-frequency mode and (d) the fundamental conjugate pair are obtained with $r=11$ and $32$ phases. The four histories in (c) nearly overlap. All panels share the elapsed-time axis; vertical dashed lines mark the arrival at $\tau=10$. The sampling interval is $0.1D/U_\infty$.}
\label{fig:forces}
\end{figure}

In the undisturbed periodic wake, lift oscillates at the fundamental shedding frequency. We therefore examine $C_L$ and its modal contributions, defined in Section~\ref{sec:modal_lift}, to relate the evolving flow structures to the forces on the cylinder.
Figure~\ref{fig:forces}(a) compares the lift histories for four phases. The arrival peaks and the subsequent lift oscillations differ across phases. Averaging the lift coefficient over all $64$ phases at each elapsed time gives the phase-ensemble mean,
\begin{equation}
 \langle C_L\rangle_\alpha(\tau)
 =\frac{1}{64}\sum_{\alpha\in\mathcal P}C_L^{(\alpha)}(\tau),
 \label{eq:lift_phase_mean}
\end{equation}
where $\mathcal P$ is the set of the $64$ phases. The phase-ensemble mean in Fig.~\ref{fig:forces}(b) changes sharply near gust arrival and continues to oscillate afterward.

We verified that $|C_L^{(\alpha)}-C_{L,\mathrm{rec}}^{(\alpha)}|<5\times10^{-3}$ at all sampled times for the $32$ phases used in the decomposition. The zero-frequency lift contribution is nearly the same for all four phases and accounts for the main change at arrival [Fig.~\ref{fig:forces}(c)].
The counterclockwise gust induces upward flow ahead of its center and downward flow behind it, consistent with the positive peak followed by the negative peak in the zero-frequency lift contribution [Fig.~\ref{fig:forces}(c)].

At arrival, the zero-frequency and fundamental lift contributions reinforce or oppose one another depending on phase. For $\alpha=\pi/2$, both contributions are positive at the positive lift peak before arrival [Fig.~\ref{fig:forces}(c,d)]. At this peak, the coefficient magnitude is close to its pre-encounter value, so the larger fundamental lift mainly reflects a change in the surface stresses of the mode. Part of this increase may arise because the fundamental mode also represents some of the common response near the cylinder. Because the tDMDpc modes are nonorthogonal, the lift attributed to each mode depends on the retained basis.

Near the subsequent negative arrival peak, the fundamental contribution opposes the zero-frequency contribution for $\alpha=0$ and $\pi/2$, and reinforces it for $\pi$ and $3\pi/2$ [Fig.~\ref{fig:forces}(d)]. The latter two cases consequently develop similar large negative peaks [Fig.~\ref{fig:forces}(a)]. During $13\leq\tau\leq20$, the contributions predominantly oppose one another for $\pi/2$ and reinforce one another for $\pi$ and $3\pi/2$. In the $\pi/2$ case, the two contributions nearly cancel around $\tau=13$, when the total lift is close to zero. The fundamental contribution remains appreciable at this time, so a small total lift does not imply a weak fundamental-mode response. The surface-stress distributions and relative timing of the two contributions determine the net force. The fundamental contribution subsequently decreases slightly.

At later times, the phase-ensemble mean continues to oscillate [Fig.~\ref{fig:forces}(b)]. The zero-frequency contribution decays [Fig.~\ref{fig:forces}(c)], and the fundamental conjugate pair becomes the main contribution to the later lift oscillations [Fig.~\ref{fig:forces}(d)]. The gust changes the amplitude and phase of the fundamental response differently across phases. Its oscillations therefore no longer cancel in the phase-ensemble mean, and the oscillations in Fig.~\ref{fig:forces}(b) are carried mainly by the fundamental conjugate pair. The phase-ensemble mean therefore does not isolate the zero-frequency contribution. tDMDpc separates the two contributions, which phase averaging cannot.
The later lift oscillations return to a common amplitude, but not to their original phase differences. This is consistent with phase-response analysis, in which the amplitude recovers while a phase shift persists\cite{TairaNakao2018}.

\section{Conclusion}
In this study, the interaction between a vortex gust and the periodic wake of a circular cylinder at $Re=100$ was investigated using a phase-controlled ensemble, tDMDpc, and a modal decomposition of the lift. The same gust was imposed at $64$ shedding phases to examine the phase dependence of the interaction. Stationary POD trajectories characterize departures from the periodic reference during the encounter, while the evolving tDMDpc modes resolve the spatial development of the localized disturbance. In the tDMDpc decomposition, the approaching gust is mainly captured by the zero-frequency mode, while gust-related structures develop in the fundamental mode during the encounter and persist near the cylinder as the gust convects downstream.

The strength of the fundamental response varies across phases, as quantified by the phase-specific coefficients. After gust arrival, the normalized coefficient magnitude is lower near $\pi/2$ and higher near $3\pi/2$. These phases show inward and outward departures of the stationary POD trajectories, respectively, during the encounter. The near-wake vorticity amplitude of the fundamental conjugate pair also decreases for $\alpha=\pi/2$ and increases for $3\pi/2$ relative to each phase's pre-encounter value over approximately $14\lesssim\tau\lesssim21$. In the fundamental-mode contributions, the counter-rotating and co-rotating configurations at $\pi/2$ and $3\pi/2$ are consistent with the attenuation and amplification observed in the POD amplitude.


The phase-ensemble mean of the lift coefficient contains oscillations from the fundamental response and therefore does not isolate the zero-frequency contribution. tDMDpc separates these contributions and relates the lift to their evolving spatial structures. The decomposition shows that the zero-frequency lift contribution accounts for most of the lift change at arrival, while the fundamental conjugate pair accounts for most of the later lift oscillations. Where both are present, they reinforce or oppose each other. For $\alpha=\pi/2$, they nearly cancel around $\tau=13$, although the fundamental contribution remains appreciable. As the wake recovers, the amplitude returns to its undisturbed value, whereas the phase shift remains.

The present results concern a single gust strength, radius, and offset. Future work should examine a wider range of gust strengths, sizes, and positions. These extensions would clarify the effects of the gust parameters on the phase-dependent vortex interactions and force contributions.

\section*{Acknowledgments}
This work was partially supported by JST SPRING, Grant Number JPMJSP2114, Japan.
This work was supported in part by the Japan Society for the Promotion of Science (JSPS) Grant-in-Aid for JSPS Fellows Grant Number JP26KJ1291. The computational resource is provided by the JAXA Supercomputer System Generation 3 (JSS3).

\clubpenalty=150
\widowpenalty=150
\displaywidowpenalty=50

\end{document}